\documentclass[aps,twocolumn,showpacs,floatfix]{revtex4}
\usepackage{amsmath}
\usepackage{amsfonts}
\usepackage{amssymb}
\usepackage{epsfig}
\usepackage{graphicx}

\begin{document}
\title{Can reactive coupling beat motional quantum limit of nano waveguides coupled to microdisk resonator}
\author{Sumei Huang and G. S. Agarwal}
\affiliation{Department of Physics, Oklahoma State University,
Stillwater, Oklahoma 74078, USA}

\date{\today}

\begin{abstract}
Dissipation is generally thought to affect the quantum nature of the system in an adverse manner, however we show that dissipatively coupled nano systems can be prepared in states which beat the standard quantum limit of the mechanical motion. We show that the reactive coupling between the waveguide and the microdisk resonator can generate the squeezing of the waveguide by injecting a quantum field and laser into the resonator through the waveguide. The waveguide can show about 70--75$\%$ of maximal squeezing for temperature about 1--10 mK. The maximum squeezing can be achieved with incident pump power of only 12 $\mu$W for a temperature of about 1 mK. Even for temperatures of 20 mK, achievable by dilution refrigerators, the maximum squeezing is about 60$\%$.
\end{abstract}
 \maketitle

\renewcommand{\thesection}{\Roman{section}}
\setcounter{section}{0}
\section{Introduction}
\renewcommand{\baselinestretch}{1}\small\normalsize

Methods for beating the standard quantum limit of radiation fields have become fairly standard. Most methods are based on nonlinear interactions of the field in a highly nonlinear medium. The question of beating the quantum limit of the mechanical motion which could range from kHz to GHz range is attracting increasing attention \cite{Mancini,Genes,Vitali,Meystre1,Meystre2,Ian,Zhou,Clerk,Woolley,Teufel,Mari,Nunnenkamp,Hertzberg}. Fortunately a nano mechanical mirror [NMO] placed in an optical cavity interacts with the field in the cavity in nonlinear fashion and this can be described by a nonlinear Hamiltonian. A scheme to beat the standard quantum limit for mechanical motion is to drive the system by a combination of a laser field and squeezed light such that the beat frequency matches the frequency of the NMO \cite{Jahne}. More recently other designs of NMO have been used \cite{Tang,Clerk1,Sankey,Anetsberger}. These have certain attractive features and appear quite versatile; for example, in the design of Li \emph{et al.} \cite{Tang}, the nano waveguide interacts reactively with the microdisk resonator. In other words the fields leak from resonator to the waveguide. Even though the coupling is of dissipative nature such a system exhibits several novel features such as normal mode splitting which traditionally was a feature of two strongly coupled oscillators described by the Hamiltonian framework \cite{Marquardt,Kippenberg,Aspelmeyer,Sumei}.

In this paper, we go one step further. We give first example of dissipative nonlinear coupling produced quantum fluctuations of the mechanical motion of the waveguide which are below the standard quantum limit. This is rather counterintuitive, as dissipation is always thought to produce negative effects, i.e., is generally thought to suppress the quantum nature of the system.

The paper is organized as follows. In Sec. II, we introduce the model, present the equation of motion for the system, and give the mean values of the system operators in steady state. In Sec. III, we calculate the quantum fluctuations in the mechanical motion of the waveguide and obtain the variance of momentum of the waveguide. In Sec. IV, we present the numerical result and show that the reactive coupling can reduce the momentum fluctuations of the waveguide below the standard quantum limit. The numbers are rather attractive; for example, at a temperature of 20 mK, achievable by a dilution refrigerator, the maximum momentum squeezing of the waveguide is about 60$\%$.
\section{Model}

Let us consider a free-standing waveguide with length $L$ interacting with a microdisk resonator \cite{Tang}. Suppose a laser with amplitude $\varepsilon_{l}$ at frequency $\omega_{l}$ drives the resonator mode $c$, and a quantum field $c_{in}$ at frequency $\omega_{s}$ is sent into the resonator through the waveguide with mass $m$ and frequency $\omega_{m}$. For convenience, we adopt the notation $Q=\sqrt{\frac{2m\omega_{m}}{\hbar}}q$ and $P=\sqrt{\frac{2}{m\hbar\omega_{m}}}p$ for the dimensionless position and momentum quadratures of the waveguide with $[Q,P]=2i$. The waveguide vibrates along the $y$ direction due to the dispersive and reactive couplings with the resonator, which are characterized by the position dependence of the resonator resonance frequency $\omega_{c}(Q)$ and the photon decay rate $\kappa_{e}(Q)$, respectively. Moreover, the waveguide is damped at a rate of $\gamma_{m}$ due to its interaction with its environment at a low temperature $T$.

In a frame rotating at the laser frequency $\omega_{l}$, the Hamiltonian describing the whole system takes the form \cite{Tang}
\begin{equation}\label{1}
\begin{array}{lcl}
\displaystyle H=\hbar
[\omega_{c}(Q)-\omega_{l}]c^{\dag}c+\frac{\hbar\omega_{m}}{4}(Q^{2}+P^{2})+\hbar\frac{
L}{c}\tilde{n}_{g}\omega_{l}\varepsilon_{l}^2\vspace{0.1in}\\\hspace{0.3in}+i\hbar\sqrt{2\kappa_{e}(Q)}[\varepsilon_{l}(c^{\dag}-c)+c^{\dag}c_{in}-c_{in}^{\dag}c].
\end{array}
\end{equation}
where the first two terms describe the free energies of the resonator and the waveguide, respectively. The third term is the interaction between the waveguide and the laser, $c$ is the speed of light in vacuum, $\tilde{n}_{g}$ is the group index
of the waveguide optical mode \cite{Pernice}, and $\varepsilon_{l}$ is related to the input power $\wp_{l}$ by
$\varepsilon_{l}=\displaystyle\sqrt{\frac{\wp_{l}}{\hbar\omega_{l}}}$. The last term gives the interactions of the resonator with the laser and the quantum field. The characteristics of the quantum field would be specified later.

For a small displacement $Q$, we can assume that both $\omega_{c}(Q)$ and $\kappa_{e}(Q)$
are coupled linearly to the displacement $Q$,
\begin{equation}\label{2}
\begin{array}{lcl}
\displaystyle\omega_{c}(Q)\approx\omega_{c}+gQ,\vspace{0.2in}\\
\displaystyle\kappa_{e}(Q)\approx\kappa_{e}+\kappa_{om}Q=\kappa_{e}(1+\eta Q),
\end{array}
\end{equation}
where $\omega_{c}$ is the resonator resonance frequency for $Q=0$, $\kappa_{e}$ is the photon decay rate for $Q=0$, and $g$ and $\kappa_{om}$ are the dispersive and reactive coupling constants between the waveguide and the resonator, respectively. We set $\displaystyle\eta=\frac{\kappa_{om}}{\kappa_{e}}$. Since in the scheme of Li \emph{et al.} \cite{Tang} the effects of reactive coupling are dominant, we will take $g\sim0$.

For simplicity, we assume that there is no intrinsic photon losses. Employing the
Heisenberg equation of motion and adding the damping and noise terms, the equations of motion for $Q$, $P$, and $c$ can be expressed
as
\begin{eqnarray}\label{3}
\displaystyle \dot{Q}&=&\omega_{m}P,\nonumber\vspace{0.2in}\\
\displaystyle \dot{P}&=&-i\eta[\tilde{\varepsilon}_{l}(c^{\dag}-c)+\sqrt{2\kappa_{e}}(c^{\dag}c_{in}-c_{in}^{\dag}c)]\nonumber\vspace{0.2in}\\& &-\omega_{m}Q-\gamma_{m}P+\xi,\nonumber\vspace{0.2in}\\
\displaystyle \dot{c}&=&-[\kappa_{e}+\kappa_{om}Q+i(\omega_{c}-\omega_{l})]c\nonumber\vspace{0.2in}\\& &+(1+\frac{\eta}{2}Q)(\tilde{\varepsilon}_{l}+\sqrt{2\kappa_{e}}c_{in}),
\end{eqnarray}
where $\tilde{\varepsilon}_{l}=\sqrt{2\kappa_{e}}\varepsilon_{l}$, and we have introduced $\xi$ as the thermal noise force acting on the waveguide with standard correlation ~\cite{Giovannetti}.
We first examine the mean values of the physical variables in steady state. These can be obtained by using the factorization ansatz i.e. mean value of the product of two operators is the same as the product of the mean values. We find that these are given by
\begin{eqnarray}\label{4}
P_{s}&=&0,\nonumber\vspace{0.2in}\\
Q_{s}&=&-\frac{2\eta}{\omega_{m}}\tilde{\varepsilon}_{l}\mbox{Im}[c_{s}],\nonumber\vspace{0.2in}\\
c_{s}&=&\frac{(1+\displaystyle\frac{\eta}{2}Q_{s})\tilde{\varepsilon}_{l}}{\kappa_{e}+\kappa_{om}Q_{s}+i\Delta},
\end{eqnarray}
where the resonator detuning $\Delta$ is defined by
\begin{equation}\label{5}
\Delta=\omega_{c}-\omega_{l}.
\end{equation}
Note that the steady-state position $Q_{s}$ of the waveguide and the steady-state complex amplitude $c_{s}$ of the resonator depend on $\eta$. In obtaining results (\ref{4}) we assumed that the quantum field $c_{in}$ had zero mean value. This would be the case generally unless the quantum field is a coherent field. We already examined the case of a coherent field in a previous publication \cite{Sumei}.

\section{BEATING THE MOTIONAL QUANTUM LIMIT FOR THE WAVEGUIDE}
In this section, we investigate whether the motional quantum limit for the waveguide can be beaten even when the basic coupling is reactive. This would be quite counterintuitive as the dissipation generally leads to the loss of decoherence and fluctuations above the quantum limit. The fluctuations in $q$ and $p$ are subject to the Heisenberg uncertainty relation. For the mechanical oscillator in ground state one has $\langle\delta q^{2}\rangle=\frac{\hbar}{2m\omega_{m}}\langle\delta Q^{2}\rangle$ and $\langle\delta p^{2}\rangle=\frac{m\hbar\omega_{m}}{2}\langle\delta P^{2}\rangle$, in which $\langle\delta Q^{2}\rangle=\langle\delta P^{2}\rangle=1$. Thus the reduction of fluctuations below unity is an indication that the standard quantum limit is broken. The question is if the fluctuations in either $Q$ or $P$ can go below the value unity.

Since we are interested in the squeezing of the waveguide, it is instructive to calculate the
fluctuations of the system's operators around their steady state values. Provided that the steady-state amplitude
of the resonator satisfies $|c_{s}|\gg1$, we linearize Eq. (\ref{3}) around its steady-state value by substituting $Q=Q_{s}+\delta Q$, $P=P_{s}+\delta
P$, and $c=c_{s}+\delta c$ into Eq. (\ref{3}), where $\delta Q$, $\delta P$, and $\delta c$ are the small fluctuations with zero
mean value. After linearization, the quantum Langevin
equations can be written in the form
\begin{equation}\label{6}
\dot{f}(t)=Zf(t)+F(t),
\end{equation}
where
\begin{equation}\label{7}
f(t)=\left(
\begin{array}{c}
\delta Q \\\delta P \\\delta c \\\delta c^{\dag}
\end{array}
\right),
\end{equation}
and $Z$ is a $4\times4$ matrix, and the quantum noise $F(t)$ is given by
\begin{equation}\label{8}
F(t)=\left(
\begin{array}{c}
0 \\ \xi-i\eta\sqrt{2\kappa_{e}}(c_{s}^{*}c_{in}-c_{in}^{\dag}c_{s}) \\ Jc_{in} \vspace{0.08in}\\Jc_{in}^{\dag}
\end{array}
\right),
\end{equation}
in which $J=\sqrt{2\kappa_{e}}(1+\frac{\eta}{2}Q_{s})$.

With the aid of the Fourier transform i.e., $f(t)=\frac{1}{2\pi}\int_{-\infty}^{+\infty}f(\omega)e^{-i\omega t}
d\omega$ and
$f^{\dag}(t)=\frac{1}{2\pi}\int_{-\infty}^{+\infty}f^{\dag}(-\omega)e^{-i\omega
t} d\omega$, where $f^{\dag}(-\omega)=[f(-\omega)]^{\dag}$, we solve Eq. (\ref{6}) in the frequency domain, and obtain the solution of Eq. (\ref{6})
\begin{equation}\label{9}
f(\omega)=VF(\omega),
\end{equation}
where $V=(-i\omega-Z)^{-1}$. From Eq. (\ref{9}), we can obtain the fluctuations in the momentum variable
\begin{eqnarray}\label{10}
 \delta P(\omega)&=&P_{T}(\omega)\xi(\omega)+P_{S}(\omega)c_{in}(\omega)+P_{S}^{*}(-\omega)c_{in}^{\dag}(-\omega),\nonumber\\
\end{eqnarray}
in which
\begin{eqnarray}\label{11}
P_{T}(\omega)&=&\frac{-i\omega}{d(\omega)}A(\omega)A^{*}(-\omega),\nonumber\\
P_{S}(\omega)&=&\eta[\frac{\omega\tilde{\varepsilon}_{l}}{d(\omega)}A(\omega)J-i\sqrt{2\kappa_{e}}c_{s}^{*}P_{T}(\omega)],
\end{eqnarray}
where
\begin{eqnarray}\label{12}
d(\omega)&=&A(\omega)A^{*}(-\omega)R-i\eta\tilde{\varepsilon}_{l}\omega_{m}[A(\omega)U-A^{*}(-\omega)U^{*}],\nonumber\\
\end{eqnarray}
and
\begin{eqnarray}\label{13}
A(\omega)&=&\kappa_{e}+\kappa_{om}Q_{s}-i(\Delta+\omega),\nonumber\\
R&=&\omega_{m}^{2}-\omega^{2}-i\gamma_{m}\omega,\nonumber\\
U&=&-\kappa_{om}c_{s}+\frac{\eta}{2}\tilde{\varepsilon}_{l}.
\end{eqnarray}
In Eq. (\ref{10}), the first term results from the thermal environment of the waveguide, the last two terms arise from the input quantum field. Thus the fluctuations in the momentum variable in the time domain would be
$\delta P(t)=\frac{1}{2\pi}\int_{-\infty}^{+\infty}\delta P(\omega)e^{-i\omega t}
d\omega$. Further the variance of momentum $\langle\delta\tilde{P}^{2}\rangle$
can be expressed as
\begin{eqnarray}\label{14}
\langle\delta
P(t)^{2}\rangle=\frac{1}{4\pi^{2}}\int\int_{-\infty}^{+\infty}
d\omega d\Omega e^{-i(\omega+\Omega)t} \langle\delta P(\omega)\delta
P(\Omega)\rangle.\nonumber\\
\end{eqnarray}
Inserting Eq. (\ref{10}) into Eq. (\ref{14}), $\langle\delta P(t)^{2}\rangle$ can be written as
\begin{eqnarray}\label{15}
\langle\delta P(t)^{2}\rangle&=&\frac{1}{4\pi^{2}}\int\int^{+\infty}_{-\infty} d\omega\,d\Omega\, e^{-i(\omega+\Omega)t}\nonumber\\
& &\{P_{T}(\omega)P_{T}(\Omega)\langle\xi(\omega)\xi(\Omega)\rangle\nonumber\\
& &+2\mbox{Re}[P_{S}(\omega)P_{S}(\Omega)\langle c_{in}(\omega)c_{in}(\Omega)\rangle]\nonumber\\
& &+P_{S}(\omega)P_{S}^{*}(-\Omega)\langle c_{in}(\omega)c^{\dag}_{in}(-\Omega)\rangle\nonumber\\
& &+P_{S}^{*}(-\omega)P_{S}(\Omega)\langle c^{\dag}_{in}(-\omega)c_{in}(\Omega)\rangle\}.
\end{eqnarray}

We assume that the quantum field is a squeezed field centered around the frequency $\omega_{s}$ with a finite width,
\begin{eqnarray}\label{16}
\langle c_{in}(\omega)
c_{in}(\Omega)\rangle=2\pi \displaystyle\frac{M\Gamma^{2}}{\Gamma^{2}+(\omega-\omega_{m})^2}\delta(\omega+\Omega-2\omega_{m}),\nonumber\\
\langle c_{in}(\omega)
c_{in}^{\dag}(-\Omega)\rangle=2\pi\left[\displaystyle\frac{N\Gamma^{2}}{\Gamma^{2}+(\omega-\omega_{m})^2}+1\right]\delta(\omega+\Omega),\nonumber\\
\end{eqnarray}
where $N=\sinh^2(r)$ and $M=\sinh(r)\cosh(r)e^{i\varphi}$ characterize the squeezed vacuum, $r$ is the
squeezing parameter of the squeezed vacuum, $\varphi$ is
the phase of the squeezed vacuum, and we set $\varphi=0$. We work in the sideband resolved limit i.e. we assume that $\omega_{s}-\omega_{l}=\omega_{m}$. The squeezed vacuum has a finite bandwidth $\Gamma$ around $\omega_{m}$, which is smaller than $\omega_{m}$ but larger than the resonator width. The antinormally ordered term has a broad band contribution coming from vacuum noise. Moreover, the thermal noise $\xi$ owns the correlation function ~\cite{Giovannetti}:
\begin{eqnarray}\label{17}
\langle\xi(\omega)\xi(\Omega)\rangle=4\pi\gamma_{m}\frac{\omega
}{\omega_{m}}\left[1+\coth\left(\frac{\hbar\omega}{2K_B T}\right)\right]\delta(\omega+\Omega),\nonumber\\
\end{eqnarray}
where $K_{B}$ is the Boltzmann constant.

Substituting Eqs. (\ref{16}) and (\ref{17}) into Eq. (\ref{15}), the time independent variance $\langle\delta P^{2}\rangle$ will be
\begin{widetext}
\begin{eqnarray}\label{18}
\langle\delta P^{2}\rangle&=&\frac{1}{2\pi}\int^{+\infty}_{-\infty}\,d\omega P_{T}(\omega)P_{T}(-\omega)2\gamma_{m}\frac{\omega}{\omega_{m}}\left[1+\coth\left(\frac{\hbar\omega}{2K_{B}T}\right)\right]\nonumber\\
& &+2\mbox{Re}\left[\frac{1}{2\pi}\int^{+\infty}_{-\infty}\,d\nu P_{S}(\omega_{m}+\nu)P_{S}(\omega_{m}-\nu)\frac{M\Gamma^{2}}{\Gamma^{2}+\nu^{2}}\right]\nonumber\\
& &+2\left[\frac{1}{2\pi}\int^{+\infty}_{-\infty} d\nu |P_{S}(\omega_{m}+\nu)|^{2}\frac{N\Gamma^{2}}{\Gamma^{2}+\nu^{2}}\right]+\frac{1}{2\pi}\int^{+\infty}_{-\infty}\,d\omega|P_{S}(\omega)|^{2}.\nonumber\\
\end{eqnarray}
\end{widetext}
The details of the calculations are given in Appendix A.

\section{Numerical results for Nano waveguide fluctuations below standard quantum limit}
We use available experimental parameters \cite{Tang}: the wavelength of the laser $\lambda=2\pi
c/\omega_l=1564.25$ nm, the mass of the waveguide $m=2$ pg (the
density of the silicon waveguide 2.33 g/cm$^3$, length 10 $\mu$m,
width 300 nm, height 300 nm),
the frequency of the waveguide $\omega_m=2\pi\times25.45$ MHz, the extrinsic photon decay rate $\kappa_{e}=0.05\omega_{m}$, the reactive coupling constant $\kappa_{om}=-2\pi\times26.6$ MHz/nm $\times\sqrt{\frac{\hbar}{2m\omega_{m}}}$, the mechanical quality factor
$Q=\omega_{m}/\gamma_{m}=5000$, and the bandwidth of the squeezed vacuum $\Gamma=5\kappa_{e}$.
\begin{figure}[htp]
 \scalebox{0.65}{\includegraphics{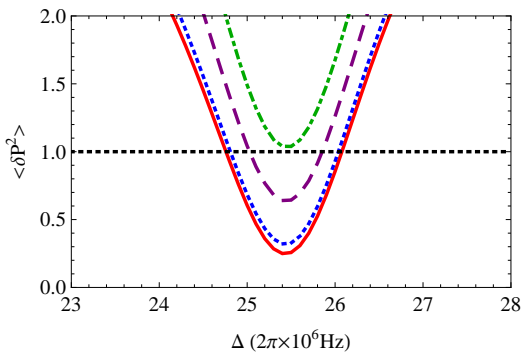}}
 \caption{\label{Fig2}(Color online)  The variance of momentum $\langle\delta P^{2}\rangle$ as a function of the detuning $\Delta$ ($2\pi\times10^{6}$Hz) for different temperatures of the environment: $T = 1$ mK (red solid), $T = 10$ mK (blue dotted), $T = 50$ mK (purple dashed), and $T = 100$ mK (green dotdashed). The horizontal dotted
line represents the standard quantum limit ($\langle\delta
P^{2}\rangle$=1). The parameters: the pump power $\wp_{l} =20$ $\mu$W, $r=1$.}
\end{figure}

We start the investigation with the influence of the reactive coupling on the squeezing of the waveguide. If the quantum field is the ordinary vacuum ($r=0$), we calculate the variances of position and momentum, and find that $\langle\delta
Q^{2}\rangle$ and $\langle\delta
P^{2}\rangle$ are always larger than unity, there is no squeezing appearance.
If the quantum field is the squeezed vacuum, and $r=1$, it has been found that there is no squeezing in the variance of position $\langle\delta Q^{2}\rangle$, but the variance of momentum $\langle\delta P^{2}\rangle$ may be squeezed.
For pump power $\wp_{l}=20$ $\mu$W, the variances of momentum  $\langle\delta
P^{2}\rangle$ versus the detuning $\Delta$ ($2\pi\times10^{6}$ Hz) for different temperatures of the environment are shown in Fig. 1. For $T=1$, 10, or 50 mK, we can see the variance of momentum $\langle\delta
P^{2}\rangle$ falls below the standard quantum limit, so the momentum squeezing takes place. The minimum value of $\langle\delta
P^{2}\rangle$ is about 0.250 at $T=1$ mK, this shows the maximum momentum squeezing of the waveguide is about 75 $\%$. Note that the maximum momentum squeezing of the waveguide decreases with increasing the temperature due to large thermal noise. Even at $T=50$ mK, the momentum squeezing is about 40$\%$.

\begin{figure}[htp]
 \scalebox{0.65}{\includegraphics{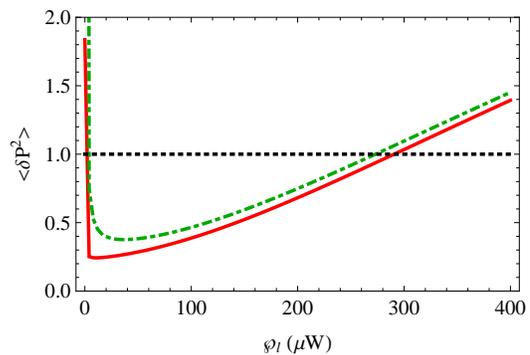}}
 \caption{\label{Fig3}(Color online)  The variance of momentum $\langle\delta P^{2}\rangle$ as a function of the pump power ($\mu$W) for different temperatures of the environment: $T = 1$ mK (red solid) and $T = 20$ mK (green dotdashed). The horizontal dotted
line represents the standard quantum limit ($\langle\delta
P^{2}\rangle$=1). The parameters: $\Delta=\omega_{m}$, $r=1$.}
\end{figure}

Next we consider the resonance case $\Delta=\omega_{m}$ in the presence of the reactive coupling, and fix $r=1$, the dependence of the variance of momentum $\langle\delta
P^{2}\rangle$ on the pump power $\wp_{l}$ ($\mu$W) for $T=1$ and 20 mK is shown in Fig. 2. It is seen that the variance of momentum $\langle\delta P^{2}\rangle$ clearly exhibits the squeezing effect over a large range of pump power ($\wp_{l}=0\sim290$ $\mu$W). The minimum value of $\langle\delta
P^{2}\rangle$ is 0.243 at a very low pump power ($\wp_{l}=12$ $\mu$W) for $T=1$ mK, so the maximum momentum squeezing of the waveguide is about 75 $\%$. For $T=20$ mK, the maximum momentum squeezing is about 60$\%$.  Note that temperatures like 20 mK are realizable by standard dilution refrigerators \cite{Press}.

\section{Conclusions}
We have shown that quantum squeezing effects in the motion of the waveguide can be generated solely due to the reactive coupling between the waveguide and the resonator by use of a squeezed vacuum. The maximum momentum squeezing is about 75$\%$, which can be achieved at a very low pump power ($\wp_{l}=12\mu$W). We show in the Appendix B the relation between the quantum fluctuations of the waveguide and the output field. Thus the squeezing of nano waveguide can be studied by examining the fluctuations of the output field of the waveguide.

\section*{ACKNOWLEDGMENT}
We gratefully acknowledge support from the NSF Grant No. PHYS
0653494.

\setcounter{equation}{0}  
\section*{APPENDIX A: THE VARIANCE OF MOMENTUM-DERIVATION OF EQUATION EQ. (18)}  
\renewcommand{\theequation}{A\arabic{equation}}
With the aid of Eq. (\ref{17}), the first term of Eq. (\ref{15}) is
\begin{eqnarray}\label{A1}
& &\frac{1}{4\pi^{2}}\int\int^{+\infty}_{-\infty} d\omega\,d\Omega\, e^{-i(\omega+\Omega)t}P_{T}(\omega)P_{T}(\Omega)\langle\xi(\omega)\xi(\Omega)\rangle\nonumber\\
&=&\frac{1}{2\pi}\int^{+\infty}_{-\infty}\,d\omega P_{T}(\omega)P_{T}(-\omega)2\gamma_{m}\frac{\omega}{\omega_{m}}\left[1+\coth\left(\frac{\hbar\omega}{2K_{B}T}\right)\right].\nonumber\\
\end{eqnarray}
Then with the help of Eq. (\ref{16}), the second term of Eq. (\ref{15}) will be
\begin{eqnarray}\label{A2}
& &\frac{1}{4\pi^{2}}\int\int^{+\infty}_{-\infty} d\omega\,d\Omega e^{-i(\omega+\Omega)t}P_{S}(\omega)P_{S}(\Omega)\langle c_{in}(\omega)c_{in}(\Omega)\rangle\nonumber\\&=&\frac{1}{2\pi}\int^{+\infty}_{-\infty} d\nu\,e^{-2i\omega_{m}t} P_{S}(\omega_{m}+\nu)P_{S}(\omega_{m}-\nu)\frac{M\Gamma^{2}}{\Gamma^{2}+\nu^{2}},\nonumber\\
\end{eqnarray}
and the third term of Eq. (\ref{15}) becomes
\begin{eqnarray}\label{A3}
& &\frac{1}{4\pi^{2}}\int\int^{+\infty}_{-\infty} d\omega\,d\Omega e^{-i(\omega+\Omega)t}P_{S}(\omega)P_{S}^{*}(-\Omega)\langle c_{in}(\omega)c^{\dag}_{in}(-\Omega)\rangle\nonumber\\
&=&\frac{1}{2\pi}\int^{+\infty}_{-\infty} d\nu |P_{S}(\omega_{m}+\nu)|^{2}\frac{N\Gamma^{2}}{\Gamma^{2}+\nu^{2}}+\frac{1}{2\pi}\int^{+\infty}_{-\infty}\,d\omega|P_{S}(\omega)|^{2}.\nonumber\\
\end{eqnarray}
Therefore, the variance $\langle\delta P(t)^{2}\rangle$ can be calculated by
\begin{widetext}
\begin{eqnarray}\label{A4}
\langle\delta P(t)^{2}\rangle&=&\frac{1}{2\pi}\int^{+\infty}_{-\infty}\,d\omega P_{T}(\omega)P_{T}(-\omega)2\gamma_{m}\frac{\omega}{\omega_{m}}\left[1+\coth\left(\frac{\hbar\omega}{2K_{B}T}\right)\right]\nonumber\\
& &+2\mbox{Re}\left[\frac{1}{2\pi}\int^{+\infty}_{-\infty}\,d\nu e^{-2i\omega_{m}t}P_{S}(\omega_{m}+\nu)P_{S}(\omega_{m}-\nu)\frac{M\Gamma^{2}}{\Gamma^{2}+\nu^{2}}\right]\nonumber\\
& &+2\left[\frac{1}{2\pi}\int^{+\infty}_{-\infty} d\nu |P_{S}(\omega_{m}+\nu)|^{2}\frac{N\Gamma^{2}}{\Gamma^{2}+\nu^{2}}\right]+\frac{1}{2\pi}\int^{+\infty}_{-\infty}\,d\omega|P_{S}(\omega)|^{2}.\nonumber\\
\end{eqnarray}
\end{widetext}
The variance has terms oscillating at twice the frequency of the nanomechanical oscillator. These terms can be removed in the standard way by working in an interaction picture defined with respect to the frequency $\omega_{m}$. This is equivalent to setting $e^{\pm2i\omega_{m}t}$ as unity, hence (A4) leads to Eq. (\ref{18}).

\setcounter{equation}{0}  
\section*{APPENDIX B: RELATION BETWEEN THE QUANTUM FLUCTUATIONS OF NANO WAVEGUIDE AND THE OUTPUT FIELD}  
\renewcommand{\theequation}{B\arabic{equation}}

In the following, we show the squeezing of the waveguide can be measured through the $y$ component of the output field.
Using the input-output relation \cite{Walls} $c_{out}(t)=\sqrt{2\kappa_{e}(Q)}c(t)$,
the fluctuations of the output field can be written as
\begin{eqnarray}\label{B1}
\delta c_{out}(\omega)&=&J\delta c(\omega)+\frac{\eta}{2}\sqrt{2\kappa_{e}}c_{s}\delta Q(\omega)\nonumber\\&=&J\delta c(\omega)+\frac{\eta}{2}\sqrt{2\kappa_{e}}c_{s}\frac{i\omega_{m}}{\omega}\delta P(\omega).\nonumber\\
\end{eqnarray}
From Eq. (\ref{9}), we find the fluctuations of the resonator field
\begin{eqnarray}\label{B2}
\delta c(\omega)&=&\frac{1}{A^{*}(-\omega)}[\frac{i\omega_{m}}{\omega}U\delta P(\omega)+Jc_{in}(\omega)].
\end{eqnarray}
Combining Eqs. (\ref{B1}) and (\ref{B2}), and defining the $y$ component of the output field as $\delta y_{out}(t)=i[\delta c_{out}^{\dag}(t)-\delta c_{out}(t)]$ so that $\delta y_{out}(\omega)=i[\delta c_{out}^{\dag}(-\omega)-\delta c_{out}(\omega)]$, one can write the fluctuations in the momentum variable of the waveguide in terms of the $y$ component of the output field
\begin{widetext}
\begin{eqnarray}\label{B3}
\delta P(\omega)&=&-\frac{\omega}{\omega_{m}}\times\frac{A(\omega)A^{*}(-\omega)\delta y_{out}(\omega)-J^{2}i[A^{*}(-\omega)c_{in}^{\dag}(-\omega)-A(\omega)c_{in}(\omega)]}{\displaystyle\frac{\eta}{2}\sqrt{2\kappa_{e}}(c_{s}^{*}-c_{s})A(\omega)A^{*}(-\omega)+J[A^{*}(-\omega)U^{*}-A(\omega)U]}.
\end{eqnarray}
\end{widetext}
It is seen that the fluctuations in the momentum variable of the waveguide is related to the $y$ component of the output field.


\begin{thebibliography}{99}

\bibitem{Mancini} S. Mancini and P. Tombesi, Phys. Rev. A \textbf{49}, 4055 (1994).
\bibitem{Genes} C. Genes, A. Mari, P. Tombesi, and D. Vitali, Phys. Rev. A \textbf{78}, 032316 (2008).
\bibitem{Vitali} D. Vitali, S. Mancini, L. Ribichini, and P. Tombesi, Phys. Rev. A \textbf{65}, 063803 (2002).
\bibitem{Meystre1} M. Bhattacharya and P. Meystre, Phys. Rev. Lett. \textbf{99}, 153603 (2007).
\bibitem{Meystre2} M. Bhattacharya and P. Meystre, Phys. Rev. Lett. \textbf{99}, 073601 (2007).
\bibitem{Ian} H. Ian, Z. R. Gong, Y. X. Liu, C. P. Sun, and F. Nori, Phys. Rev. A \textbf{78}, 013824 (2008).
\bibitem{Zhou} X. Zhou and A. Mizel, Phys. Rev. Lett. \textbf{97}, 267201 (2006).
\bibitem{Clerk} A. A. Clerk, F. Marquardt, and K. Jacobs, New J. Phys. \textbf{10}, 095010 (2008).
\bibitem{Woolley} M. J. Woolley, A. C. Doherty, G. J. Milburn, and K. C. Schwab, Phys. Rev. A \textbf{78}, 062303 (2008).
\bibitem{Teufel} J. D. Teufel, T. Donner, M. A. Castellanos-Beltran, J. W. Harlow, and K. W. Lehnert,
Nature Nanotechnology, \textbf{4}, 820(2009).
\bibitem{Mari} A. Mari and J. Eisert, Phys. Rev. Lett. \textbf{103}, 213603 (2009).
\bibitem{Nunnenkamp} A. Nunnenkamp, K. Borkje, J. G. E. Harris, and S. M. Girvin, arXiv:1004.2510v2.
\bibitem{Hertzberg} J. B. Hertzberg, T. Rocheleau, T. Ndukum, M. Savva, A. A. Clerk, and K. C. Schwab, Nature Physics \textbf{6}, 213 (2010).
\bibitem{Jahne} K. J\"{a}hne, C. Genes, K. Hammerer, M. Wallquist, E. S. Polzik, and P. Zoller, Phys. Rev. A \textbf{79}, 063819 (2009); S. Huang and G. S. Agarwal, arXiv:0905.4234.
\bibitem{Tang} M. Li, W. H. P. Pernice, and H. X. Tang, Phys. Rev. Lett. {\bf 103}, 223901
(2009).
\bibitem{Clerk1} F. Elste, S. M. Girvin, and A. A. Clerk, Phys. Rev. Lett. {\bf 102}, 207209 (2009).
\bibitem{Sankey} J. C. Sankey, C. Yang, B. M. Zwickl, A. M. Jayich, and J. G. E. Harris,  Nature Phys. {\bf 6}, 707 (2010). 
\bibitem{Anetsberger}  G. Anetsberger, E. Gavartin, O. Arcizet, Q. P. Unterreithmeier, E. M. Weig, M. L. Gorodetsky, J. P. Kotthaus, and T. J. Kippenberg, arXiv:1003.3752v1.
\bibitem{Marquardt} F. Marquardt, J. P. Chen, A. A. Clerk, and S. M. Girvin, Phys. Rev. Lett. {\bf 99}, 093902 (2007).
\bibitem{Kippenberg} J. M. Dobrindt, I. Wilson-Rae, and T. J. Kippenberg, Phys. Rev. Lett. {\bf 101}, 263602 (2008).
\bibitem{Aspelmeyer} S. Gr\"{o}blacher, K. Hammerer, M. Vanner, and M. Aspelmeyer, Nature (London) {\bf 460}, 724 (2009).
\bibitem{Sumei} S. Huang and G. S. Agarwal, Phys. Rev. A {\bf 81}, 053810 (2010).

\bibitem{Pernice} W. H. P. Pernice, M. Li, and H. X. Tang, Opt. Express {\bf 17}, 1806 (2009).

\bibitem{Giovannetti} V. Giovannetti and D. Vitali, Phys. Rev. A \textbf{63}, 023812 (2001).
\bibitem{Press} Press release, 28th of January 2010, Oxford instruments, www.oxford-instruments.com; M. J. Mart\'{i}nez-P\'{e}rez, J. Ses\'{e}, F. Luis, D. Drung, and T. Schurig, Rev. Sci. Instrum. \textbf{81}, 016108 (2010).
\bibitem{Walls} D. F. Walls and G. J. Milburn, \textit{Quantum Optics} (Springer-Verlag, Berlin, 1994).
\end{thebibliography}
\end{document}